\PassOptionsToPackage{table}{xcolor}
\documentclass[aps,prd,twocolumn,superscriptaddress,nofootinbib,showpacs,floatfix]{revtex4-2}
\usepackage{adjustbox}
\usepackage[utf8]{inputenc}
\usepackage{graphicx,epsfig}
\usepackage{amsmath,amssymb,dsfont,epsfig,graphicx,xcolor,fontenc,physics}
\usepackage{bm}
\usepackage[english]{babel}
\usepackage{graphicx}%
\usepackage{amsfonts}%
\usepackage{amssymb}
\usepackage{braket}
\usepackage{hyperref}
\usepackage{epstopdf}
\usepackage{ulem}
\usepackage{soul}
\usepackage{float}
\usepackage{booktabs}
\usepackage{multirow}
\usepackage{cancel}
\definecolor{nicered}{rgb}{0.7,0.1,0.1}
\definecolor{nicegreen}{rgb}{0.1,0.5,0.1}
\hypersetup{colorlinks,citecolor= nicegreen,linkcolor= nicered}

\newcommand{\dgm}{\delta g_M}
\newcommand{\keV}{\,\text{keV}}

\emergencystretch=\maxdimen
\def\IFIC{Instituto de F\'isica Corpuscular,
CSIC - Universitat de Val\'encia, Spain 
}
\def\IFAE{Institut de F\'isica d'Altes Energies (IFAE), The Barcelona Institute of Science and Technology, Campus UAB, 08193 Cerdanyola del Vall\`es (Barcelona), Spain}

\begin{document}

\title{First observation of weak magnetism in neutron decay}

\author{Miriam Camarasa-Domene}
\email[Electronic address: ]{miriam.camarasa@ific.uv.es}
\affiliation{\IFIC}

\author{Mart\'in Gonz\'alez-Alonso}
\email[Electronic address: ]{martin.gonzalez@ific.uv.es}
\affiliation{\IFIC}
\author{Mart\'in Novoa-Brunet}
\email[Electronic address: ]{mnovoa@ifae.es}
\affiliation{\IFIC}
\affiliation{\IFAE}

\begin{abstract}
Weak magnetism, introduced by Gell-Mann as a hallmark consequence of the
conserved vector current (CVC) hypothesis, has remained unobserved in the decay of
the free neutron, the simplest $\beta$-decay system. The reason is that the relevant
observable, the $\beta$ asymmetry, cannot separate the
weak-magnetism form factor $g_M$ from the axial-to-vector ratio $\lambda$. We show
that this degeneracy is lifted by combining the $\beta$ asymmetry measurements with
the neutron lifetime and the $\mathcal{F}t$ value of superallowed $0^+\to0^+$ decays.
Current data yield $g_M = 4.47(87)$, in agreement with the CVC prediction of
$4.71$ and excluding the absence of weak magnetism at the $5.1\sigma$ level. 
Importantly, the significance remains between $5.0\sigma$ and $6.1\sigma$ for all considered variations of the neutron dataset, despite its well-known internal tensions.
We further show that, under present experimental
conditions, a full analysis of the energy dependence of the asymmetry would improve
the $g_M$ extraction by only a few percent. These results constitute the first observation of
weak magnetism in the decay of the free neutron nearly seven decades after Gell-Mann's prediction. 
\end{abstract}

\maketitle

%%%%%%%%%%%%%%%%%%%%%%%%%%%%%%%%%%%%%%%%%%
\section{Introduction}
\label{sec:introALT}
%%%%%%%%%%%%%%%%%%%%%%%%%%%%%%%%%%%%%%%%%%

Precision studies of $\beta$ decays have played a pivotal role in shaping our understanding of the weak interaction. From the discovery of the neutrino and parity violation to establishing the $V\!-\!A$ structure of the Standard Model (SM), these processes continue to provide stringent tests of fundamental physics~\cite{Vos:2015eba,Gonzalez-Alonso:2018omy}. Central to this framework is the conserved vector current (CVC) hypothesis~\cite{Gershtein:1955fb,Feynman:1958ty}, which now arises naturally from the SM gauge structure: it relates the weak vector current to the isovector electromagnetic current as members of the same isospin triplet, thereby linking the weak-vector form factors to their electromagnetic counterparts.

A cornerstone prediction of CVC is weak magnetism (WM)~\cite{Gell-Mann:1958sai}. Just as the matrix element of the electromagnetic current features a Pauli term ($\sigma^{\mu\nu} q_\nu$) representing the anomalous magnetic coupling, the weak vector current generates an identical tensorial structure whose form factor is directly related by CVC to electromagnetic observables -- the so-called strong form of the hypothesis. 
Gell-Mann proposed testing it 
by comparing the spectral shapes of the mirror $\beta$ decays of $^{12}\text{B}$ and $^{12}\text{N}$ with the strength of the analog $M1$ $\gamma$ transition in $^{12}\text{C}$. This landmark test was carried out by Lee, Mo, and Wu~\cite{Lee:1963zz}, providing important early evidence for CVC. WM has since been observed in some nuclear $\beta$ transitions, though with varying reliance on nuclear-structure input~\cite{Severijns:2006dr}.

In contrast, weak magnetism has never been directly measured in the decay of the free neutron, the simplest $\beta$-decay process. This gap is particularly notable because WM is the \textit{only} non-negligible form factor beyond the leading $V\!-\!A$ couplings that contributes to the differential decay rate at first order in recoil. Furthermore, experimental precision in neutron observables has advanced by two to three orders of magnitude since Gell-Mann's paper, with remarkable progress in both lifetime and correlation measurements~\cite{Abele:2008zz,Dubbers:2021wqv,ParticleDataGroup:2024cfk}.

To formalize the concept of WM in neutron decay, we consider the matrix element of the weak current, which can be decomposed as:
\begin{align}
    \langle p|\,\bar{u}\gamma^\mu d\,|n\rangle &= \phantom{-}\bar{u}_p\bigg[ g_V\,\gamma^\mu - \frac{g_M-g_V}{2m_N}\,i\sigma^{\mu\nu}q_\nu \bigg] u_n\,, \nonumber\\
    \langle p|\,\bar{u}\gamma^\mu\gamma_5 d\,|n\rangle &= -\bar{u}_p\bigg[ g_A\,\gamma^\mu\gamma_5 \bigg] u_n\,, \label{eq:FFdecompositionVA}
\end{align}
where $u_{n(p)}$ is the neutron (proton) spinor, $q \equiv p_n - p_p$ is the momentum transfer, and $m_N \equiv (m_n + m_p)/2$ is the average nucleon mass. Here, $g_V$, $g_A$, and $g_M$ denote the form factors at $q^2=0$. Their $q^2$ dependence, along with the contributions from the induced (pseudo)scalar and tensor form factors, can be safely neglected due to recoil and isospin suppression~\cite{Weinberg:1958ut}. 

Under CVC, $g_V = 1$ up to second-order isospin-breaking corrections~\cite{Behrends:1960nf,Ademollo:1964sr,Donoghue:1990ti}\footnote{We set $g_V = 1$ hereafter. This is merely a normalization choice, as our analysis is sensitive only to the ratios $g_A/g_V$ and $g_M/g_V$.}, and the WM coupling $g_M$ is linked to the magnetic moments of the proton and neutron, yielding~\cite{ParticleDataGroup:2024cfk}:
\begin{equation}
    g_M^{\text{CVC}} 
    = \frac{\mu_p-\mu_n}{\mu_N} \,=\, 4.71\,,
    \label{eq:gM_CVC}
\end{equation}
with a negligible uncertainty for the purposes of this work and where $\mu_N$ is the nuclear magneton $e/(2m_p)$.

Despite massive advances in neutron decay experiments, extracting WM has faced a persistent impasse. The most precise observable, the lifetime $\tau_n$, is essentially insensitive to $g_M$~\cite{Weinberg:1959zza,Wilkinson:1982hu}. 
The $\beta$ spectrum does depend on it, but would not reach the needed precision, due to current systematic uncertainties and the narrow energy range available~\cite{Hickerson:2017fzz}.
The most precise correlation coefficient, the $\beta$ asymmetry~\cite{Bopp:1986rt,Liaud:1997vu,Erozolimsky:1997wi,Mund:2012fq,UCNA:2017obv,Markisch:2018ndu}, has sufficient WM sensitivity (the effect shifts the asymmetry by roughly $1.4\%$, eight times the PERKEO~III precision~\cite{Markisch:2018ndu}) but cannot disentangle $g_M$
from the $g_A/g_V$ ratio: a single measurement cannot determine both parameters simultaneously.

For this reason, past experimental analyses have fixed the weak-magnetism contribution to its CVC value and used the $\beta$ asymmetry to extract $g_A/g_V$.
More precisely, one extracts $\lambda\equiv \frac{g_A}{g_V}\left(\frac{1+\Delta_R^A}{1+\Delta_R^V}\right)^{\!1/2}$, which absorbs the inner radiative corrections that accompany the $g_A/g_V$ ratio in all observables~\cite{Gonzalez-Alonso:2018omy}.

As a result, weak magnetism has never been established in neutron decay, the cleanest and most fundamental of all $\beta$ transitions. In this Letter, we address this challenge through a combined data analysis: by leveraging the $\mathcal{F}t^{0^+\to 0^+}$ input
alongside the neutron lifetime to precisely pin down $\lambda$, our goal is to break the parameter degeneracy and isolate the WM contribution from the experimental $\beta$ asymmetry.

%%%%%%%%%%%%%%%%%%%%%%%%%%%%%%%%%%%%%%%%%%
\section{Methodology and input}
\label{sec:formalism}
%%%%%%%%%%%%%%%%%%%%%%%%%%%%%%%%%%%%%%%%%%

Our starting point is the relation~\cite{Czarnecki:2018okw}
\begin{equation}
    r_n\,\tau_n(1+3\lambda^2)= 2\,\mathcal{F}t^{0^+\to0^+}\,,
    \label{eq:lifetime_kappa}
\end{equation}
which allows us to extract $\lambda$ from the neutron lifetime, $\tau_n$, and the superallowed $\mathcal{F}t$ value (we drop the superscript $0^+\to0^+$). Here $r_n \equiv f_n(1+\delta_R^{\prime(n)})\ln 2 =1.18796(7)$, with $f_n = 1.6887(1)$~\cite{Czarnecki:2018okw} the neutron phase-space factor and $\delta_R^{\prime(n)} = 1.4902(2)\%$~\cite{Towner:2010zz} the long-distance radiative correction to the neutron decay rate. The uncertainty on $r_n$ is negligible for this work and will be omitted hereafter. 

It is instructive to linearize Eq.~\eqref{eq:lifetime_kappa} around a set of reference values $(\tau_{n,0}, \mathcal{F}t_0, \lambda_0)$ satisfying it. This yields
\begin{equation}
    \delta\lambda = \frac{1 + 3\lambda_0^2}{6\lambda_0} \left( \frac{\delta\mathcal{F}t}{\mathcal{F}t_0} - \frac{\delta\tau_n}{\tau_{n,0}} \right)~,
    \label{eq:deltalambda}
\end{equation}
with $\delta X \equiv X - X_0$ being the variation relative to the reference value. Hence the relative uncertainties of  $\tau_n$ and ${\cal F}t$ enter the determination of $\lambda$ with equal weight, as expected from Eq.~\eqref{eq:lifetime_kappa}.

Superallowed transitions between $J^P=0^+$, $T=1$ isobaric analog states proceed exclusively through the vector current. For each such transition, the measured $ft$ value (the product of the statistical rate function $f$ and the partial half-life $t$) is corrected for radiative and isospin-symmetry-breaking effects to give $\mathcal{F}t\equiv ft\,(1+\delta_R')(1+\delta_{\rm NS}-\delta_C)$, the nucleus-independent quantity entering Eq.~\eqref{eq:lifetime_kappa}. 
This corrected value is typically extracted by averaging the most precisely measured  transitions, and used for the determination of $V_{ud}$~\cite{Hardy:2020qwl,Gorchtein:2023naa}. 
However, finite-size corrections to the ${\cal F}t$ value were recently shown to depend on the parent charge radius more strongly than assumed~\cite{Seng:2023cgl,Gorchtein:2025wli}. So far this has been accounted for only in $^{26m}$Al, the most precisely determined transition, reanalyzed with updated charge radii~\cite{Gorchtein:2025wli}, yielding
\begin{equation}
    \mathcal{F}t 
    = 3070.01\,(1.90)\,{\rm s}~,
    \label{eq:Ft_Al}
\end{equation}
where the uncertainty combines in quadrature $1.23\,{\rm s}$
(statistical), $0.20\,{\rm s}$ ($\delta_R'$), and $1.44\,{\rm s}$ ($\delta_{\rm NS}$)~\cite{Gorchtein:2025wli}. We adopt this value as our
baseline input in what follows.

The neutron lifetime, $\tau_n$, is the most precisely known neutron-decay observable. 
Modern determinations rely on two complementary techniques~\cite{Wietfeldt:2011suo}. In \textit{beam} experiments, the decay rate is measured by counting emitted protons, whereas \textit{bottle} experiments count the surviving neutrons stored in material or magnetic traps. However, the two methods are in a long-standing tension exceeding $4\sigma$. Following the PDG selection~\cite{ParticleDataGroup:2024cfk}, which retains eight bottle measurements and discards the beam values, we combine them according to the PDG prescription, which yields
\begin{equation}
    \tau_n = 878.36(43)\,{\rm s}\,,
    \label{eq:tau_avg}
\end{equation}
where the uncertainty includes a scale factor $S=1.81$ due to internal tension within the bottle dataset itself. 

Using Eqs.~\eqref{eq:Ft_Al} and \eqref{eq:tau_avg} in Eq.~\eqref{eq:lifetime_kappa} yields
\begin{align}
\lambda&=-1.27598(61)\,,
\label{eq:lambdaLifetime}
\end{align}
whose central value we adopt as the reference point $\lambda_0$.\footnote{Alternatively, $\lambda$ can be {\it calculated} from lattice-QCD $g_A/g_V$ and $\Delta_R^A-\Delta_R^V$, but current results~\cite{FlavourLatticeAveragingGroupFLAG:2024oxs,Chang:2018uxx,Jang:2023zts,Alexandrou:2023qbg,Cirigliano:2022hob,Ma:2023kfr} are an order of magnitude less precise than Eq.~\eqref{eq:lambdaLifetime}.}

Let us now discuss the $\beta$ asymmetry correlation. At leading order, the electron angular distribution in the decay of polarized neutrons takes the form 
\begin{equation}
    d\Gamma \propto 1 + \frac{p_e}{E_e} P A_{\rm LO}\cos\theta~,
    \label{eq:LOdistribution}
\end{equation}
where $P$ is the neutron polarization, $\theta$ the angle between the neutron spin direction and the electron momentum, and $A_{\rm LO} = -2\,\lambda(\lambda+1)/(1+3\lambda^2)$. 
Weak magnetism, as well as other recoil and radiative corrections, introduces additional terms in the angular distribution. They are typically removed by the experimental collaborations, which translate their measurements into an extraction of $A_{\rm LO}$. This implicitly assumes SM physics and CVC for the WM term. To avoid the CVC assumption, we instead reinterpret each measurement as an effective asymmetry $\tilde A$, reintroducing the dependence on $g_M$ and $\lambda$ due to the recoil corrections~\cite{Falkowski:2021vdg}.

Before detailing this reinterpretation, let us present the current status of the
measurements of $\tilde A$. 
The available dataset~\cite{Bopp:1986rt,Liaud:1997vu,Erozolimsky:1997wi,Mund:2012fq,UCNA:2017obv,Markisch:2018ndu} is
clearly dominated by the PERKEO III measurement, $\tilde{A} =
-0.11985(21)$~\cite{Markisch:2018ndu}. For our baseline input, we use again the PDG
average~\cite{ParticleDataGroup:2024cfk}:
\begin{equation}
\tilde{A} = -0.11958(21)~,
\label{eq:Atilde_avg}
\end{equation}
which includes a scale factor of $S=1.15$, reflecting a well-known internal tension between the modern~\cite{Mund:2012fq,UCNA:2017obv,Markisch:2018ndu} and the older~\cite{Bopp:1986rt,Liaud:1997vu,Erozolimsky:1997wi}
measurements.  

We now turn to the WM corrections and how to incorporate the $g_M$ sensitivity of $\tilde A$ in our analysis. At first order in recoil, the WM form factor modifies the angular distribution as~\cite{Falkowski:2021vdg} (see also~\cite{Holstein:1974zf,Wilkinson:1982hu,Ando:2004rk})
\begin{equation}
    d\Gamma \propto 1 + \xi_b + \frac{p_e}{E_e} P \left(A_{\rm LO} + \Delta A\right)\cos\theta\,,
    \label{eq:dW_recoil}
\end{equation}
where $\xi_b$ and $\Delta A$ are energy-dependent corrections, 
whose explicit expressions in terms of
$g_M$ and $\lambda$ are given in Appendix~\ref{app:formulas}. For our purposes, it suffices
to note that $\xi_b$ involves corrections proportional to
$m_e/E_e$ and $E_e$, while $\Delta A$ involves a term
proportional to $E_e$. Other recoil-order terms, as well as the outer radiative corrections, also modify the distribution~\cite{Shann:1971fz,Holstein:1974zf,Wilkinson:1982hu,Ando:2004rk,Falkowski:2021vdg}, but since they are fully known, independent of $g_M$ and insensitive to small variations of $\lambda$, we do not discuss them further.

The calculation of the effective asymmetry $\tilde A$ as a function of $g_M$ and $\lambda$ requires integrating $\xi_b(E_e)$ and $\Delta A(E_e)$ over the measured $E_e$ window, which introduces the energy averages $\langle m_e/E_e\rangle$, $\langle E_e\rangle$, and $\langle E_e\rangle_p$ (the latter weighted by an extra $p_e/E_e$ factor).
Since these averages depend on the $E_e$ window, they vary from experiment to experiment. 
 As a simplification, we first perform a weighted average across experiments (Table~\ref{tab:beta_decay_A0_moments_cAi_summary}), and treat the resulting $\tilde A$ value in Eq.~\eqref{eq:Atilde_avg} as a single effective observable.

\begin{table}[t]
\begin{tabular}{lccccc}
\toprule
Dataset &  $\tilde{A}$ & $\langle E_e\rangle$  & $\langle E_e\rangle_p$ & $\big\langle \frac{m_e}{E_e}\big\rangle$ & ~$c_A^{\text{wm}}\!\times\! 10^{3}$ \\
\midrule
All $\tilde A$ & ~$-0.11958(21)$~  & 0.944 & 0.948 & 0.548 &  2.97 \\
Post-2012 $\tilde A$~ & ~$-0.11979(19)$~ & 0.945 & 0.950 & 0.547 & 2.98 \\
Ref.~\cite{Falkowski:2021vdg}  & ~$-0.11958(21)$~ & 0.812 & 0.841 & 0.569 &  2.52 \\
\bottomrule
\end{tabular}
\caption{Weighted averages of $\tilde{A}$ and the energy moments for different datasets, and the corresponding $c_A^{\text{wm}}$ coefficient at the reference value $\lambda_0$. $\langle E_e\rangle$ and $\langle E_e\rangle_p$ are given in MeV.
``All'' refers to ~\cite{Bopp:1986rt,Liaud:1997vu,Erozolimsky:1997wi,Mund:2012fq,UCNA:2017obv,Markisch:2018ndu}, and ``Post-2012'' to ~\cite{Mund:2012fq,UCNA:2017obv,Markisch:2018ndu}.}
\label{tab:beta_decay_A0_moments_cAi_summary}
\end{table}

Explicit expressions for $\xi_b$, $\Delta A$, the energy averages above, and the resulting $\tilde A(\lambda,g_M)$ are collected in Appendix~\ref{app:formulas}. Here it is more instructive to show the following linearized result:
\begin{equation}
    \frac{\delta\tilde A}{\tilde A_0} = -3.11\,\delta\lambda 
    + c_A^{\text{wm}}\,\delta g_M\,,
    \label{eq:Atilde_linear}
\end{equation}
where $\delta\tilde A$ and $\delta g_M$ are the variations with respect to the reference values $\tilde A_0 \equiv A_{\rm LO}(\lambda_0) =-0.11969$ and $g_M^{\text{CVC}}$. 
The $c_A^{\text{wm}}$ coefficient contains the dependence on the energy averages discussed above, namely \mbox{$10^3c_A^{\text{wm}}= -1.42 - 0.92 \langle E_e\rangle + 0.24 \langle m_e/E_e\rangle + 5.42 \langle E_e\rangle_p$} with $\langle E_e\rangle$ and $\langle E_e\rangle_{p}$ in MeV. Using the values in Table~\ref{tab:beta_decay_A0_moments_cAi_summary}, this gives $c_A^{\text{wm}}=2.97\times 10^{-3}$. 
The smallness of this factor reflects the recoil suppression of the WM contribution: a per-mil (relative) measurement of $\tilde{A}$ translates into an absolute $g_M$ uncertainty of only $\sim0.3$ (plus the contribution from $\lambda$).

Combining Eqs.~\eqref{eq:deltalambda} and \eqref{eq:Atilde_linear} yields the WM coupling
\begin{equation}
    g_M 
    = g_M^{\text{CVC}} - 804\,\left(\frac{\delta\mathcal{F}t}{\mathcal{F}t_0} - \frac{\delta\tau_n}{\tau_{n,0}} \right)+337\,\frac{\delta\tilde{A}}{\tilde{A}_0}~,
    \label{eq:deltagMNum}
\end{equation}
which is one of our main results. Recall that $\delta X\equiv X-X_0$, with the reference values $\mathcal{F}t_0 =3070.01\,{\rm s}$, $\tau_{n,0}=878.36\,{\rm s}$ and ${\tilde A}_0=-0.11969$.

%%%%%%%%%%%%%%%%%%%%%%%%%%%%%%%%%%%%%%%%%%
\section{Results}
\label{sec:results}
%%%%%%%%%%%%%%%%%%%%%%%%%%%%%%%%%%%%%%%%%%

Our relation, Eq.~\eqref{eq:deltagMNum}, expresses $g_M$ directly in terms of the superallowed $\mathcal{F}t$ value, the neutron lifetime $\tau_n$, and the $\beta$ asymmetry $\tilde A$. Inserting the current experimental values for these three inputs, discussed above and summarized in Eqs.~\eqref{eq:Ft_Al},~\eqref{eq:tau_avg} and~\eqref{eq:Atilde_avg}, gives $g_M = 4.40(50)_{\mathcal{F}t}(39)_{\tau_n}(59)_{\tilde A}=4.40(87)$, showing that the
three inputs contribute comparably to the total uncertainty.

This simple estimate relies on a few approximations, introduced to keep the discussion transparent and isolate the essential ingredients. At this point, however, a more refined treatment is warranted: a global fit built from the full (rather than linearized) expressions, retaining the small WM contribution to the neutron lifetime itself~\cite{Weinberg:1959zza,Wilkinson:1982hu}, and treating each $\beta$ asymmetry measurement individually rather than through their PDG average. Only this last point has a small but visible effect, and is worth elaborating on: since each experiment covers a different electron energy window, the averages $\langle E_e\rangle$, $\langle E_e\rangle_p$, and $\langle m_e/E_e\rangle$ differ slightly from one measurement to the next, and so does the $c_A^{\text{wm}}$ coefficient multiplying $g_M$ in Eq.~\eqref{eq:Atilde_linear}. 

\begin{figure}
    \centering
    \includegraphics[width=0.95\linewidth]{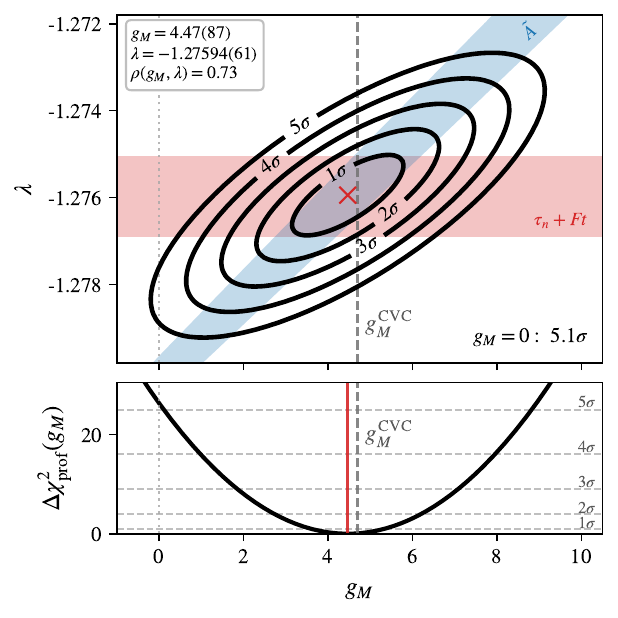}\hspace{0.05\linewidth}
    \caption{Two-dimensional and profiled constraints on $\lambda$ and $g_M$ from the combined analysis of $\tau_n$, $\mathcal{F}t$, and the $\beta$ asymmetry measurements included in the PDG analysis. 
    Colored shaded regions are the $\Delta\chi^2=2.3$ regions from the individual observable classes.}
    \label{fig:contour_baseline}
\end{figure}

The result of the corresponding global $\chi^2$ minimization is shown in Fig.~\ref{fig:contour_baseline}, along with the profiled $\Delta\chi^2(g_M)$, from which we obtain the main result of this Letter:
\begin{equation}
    g_M = 4.47(87)~.
    \label{eq:gM_result}
\end{equation}
The absence of weak magnetism, $g_M=0$, is excluded at the $5.1\sigma$ level, while the extracted value agrees with the CVC prediction of Eq.~\eqref{eq:gM_CVC} at the $0.3\sigma$ level. This constitutes the first observation of the weak-magnetism form factor in neutron decay.

Given the existing tension within the various neutron-decay measurements, it is important to assess the robustness of the extraction against different dataset choices. 
This assessment is summarized in Table~\ref{tab:gM_scenarios}. We observe that restricting the asymmetry input to the three post-2012 measurements removes the need for an ad hoc scale factor and increases the significance to $6.1\sigma$. Using only the most precise single measurement of each observable, PERKEO\,III~\cite{Markisch:2018ndu} and UCN$\tau$~\cite{Musedinovic:2024gms}, yields $g_M = 4.68(81)$, which lies $5.8\sigma$ away from zero. On the lifetime side, enlarging the dataset with the measurements excluded by the PDG (the beam values and one bottle measurement~\cite{Yue:2013qrc,Byrne:1996zz,Mampe:1993an}) gives $\tau_n = 878.52(54)\,{\rm s}$, 
with an inflated scale factor, $S=2.3$, yielding $g_M=4.63(92)$ and
thus almost no impact on the final significance. 
Finally, the same happens for a global fit that also includes the remaining correlation coefficients ($a$, $B$, $\lambda_{AB}$) ~\cite{Schumann:2007B,Beck:2024aspect,Wietfeldt:2024acorn,Byrne:2002tx,Stratowa:1978gq,Grigoriev:1968zz,Kreuz:2005jz,Serebrov:1998aj,Kuznetsov:1995sk,Christensen:1970hg,Erozolimsky:1970zz,Mostovoi:2001ye} following the prescriptions described in Appendix~\ref{app:formulas}.
The small shift in the central value is driven mainly by the additional constraint on $\lambda$ from the $a$ measurements, particularly aSPECT~\cite{Beck:2024aspect}. Despite their lower precision, tension among these additional observables produces a non-negligible pull on the preferred value, while their direct sensitivity to WM plays only a minor role.

Overall, Table~\ref{tab:gM_scenarios} demonstrates that our extraction in Eq.~\eqref{eq:gM_result} is robust against the dataset selection and conservative, as higher significances are easily obtained with reasonable choices. In all cases the extracted value is compatible with the CVC expectation within $1\sigma$, and the absence of WM is excluded with a significance ranging from $5.0\sigma$ to $6.1\sigma$.

\begin{table}[t]
  \centering

  \begin{tabular}{lcc}
    \toprule
    Neutron dataset & $g_M$ & Signif. \\
    \midrule
    {\bf Baseline} $\boldsymbol{(\tilde{A}_{\text{\bf PDG}} + \tau_{n,\text{\bf PDG}} )}$\hspace{0.7cm} & $\boldsymbol{4.47(87)}$ & $\boldsymbol{5.1\sigma}$ \\
    $\tilde{A}_{\text{post-2012}}$ + $\tau_{n,\text{PDG}}$   & $5.00(82)$ &  $6.1\sigma$ \\
    $\tilde{A}_{\text{best}}$ + $\tau_{n,\text{best}}$  & $4.68(81)$ & $5.8\sigma$ \\
    $\tilde{A}_{\text{PDG}}$ + $\tau_{n,\text{all}}$  & $4.63(92)$ & $5.0\sigma$ \\
    $\tilde{A}_{\text{PDG}}$ + $\tau_{n,\text{PDG}}$ + ($a, B, \lambda_{AB})$
    & $4.85(85)$ & $5.7\sigma$ \\
    \bottomrule
  \end{tabular}
  \caption{Weak-magnetism extraction for different neutron datasets. The first row is the  baseline analysis; the following rows modify it as indicated. See main text for details.
  The last column gives the significance of the exclusion of $g_M=0$. 
  }
  \label{tab:gM_scenarios}
\end{table}

Our analysis builds upon the work of Ref.~\cite{Falkowski:2021vdg}, whose global fit found evidence for nucleon WM at the $3\sigma$ level, and improves on it in four aspects, elaborated below.

First, rather than relying only on a global fit whose internal dependencies can be difficult to disentangle, our work includes a transparent, stepwise procedure that makes explicit the underlying logic of the extraction and identifies the primary driving observables. This is directly embodied in Eq.~\eqref{eq:deltagMNum}, which expresses $g_M$ as a function of $\mathcal{F}t$, $\tau_n$, and $\tilde{A}$.

Second, we update the experimental and theoretical inputs to incorporate recent developments: specifically, the superallowed $\mathcal{F}t$ value from Ref.~\cite{Gorchtein:2025wli}, and an updated lifetime dataset that incorporates the latest UCN$\tau$ measurement~\cite{Musedinovic:2024gms}.

Third, we refine the theoretical description of the $\beta$ asymmetry $\tilde{A}$ as a function of $g_M$ and $\lambda$.
In particular, we calculate the energy averages discussed in the previous section taking into account the actual experimental energy window, which is substantially narrower for all experiments than the full kinematically allowed range.
In contrast, Ref.~\cite{Falkowski:2021vdg} integrated over the entire physical phase space for $\langle E_e \rangle$ and $\langle E_e \rangle_p$,\footnote{For $\langle m_e/E_e \rangle$, Ref.~\cite{Falkowski:2021vdg} adopted the value $\langle m_e/E_e \rangle = 0.569$~\cite{Gonzalez-Alonso:2018omy}, which corresponds to the (unweighted) average of the values obtained integrating over the  experimental window of each measurement, cf. Table 7 in Ref.~\cite{Gonzalez-Alonso:2018omy}.} which underestimates the sensitivity to WM, cf.~Table~\ref{tab:beta_decay_A0_moments_cAi_summary}. 

Finally, we systematically investigate the robustness of the extraction against different dataset selections, an essential step given the known tensions in the neutron data. 

Taken together, these refinements yield a more precise and robust extraction of WM, establishing a significance consistently above the $5\sigma$ threshold, irrespective of the dataset choice.

%%%%%%%%%%%%%%%%%%%%%%%%%%%%%%%%%%%%%%%%%%
\section{Spectral information}
\label{sec:shape_main}
%%%%%%%%%%%%%%%%%%%%%%%%%%%%%%%%%%%%%%%%%%

We also go further than Ref.~\cite{Falkowski:2021vdg} by examining the energy dependence of the asymmetry. The extraction presented above uses only the effective, energy-averaged value $\tilde A$.  Since the asymmetry is measured in energy bins, a natural question arises: does its spectral shape carry additional information on $g_M$ that could significantly improve its extraction?

\begin{figure}
    \centering
    \includegraphics[width=\linewidth]{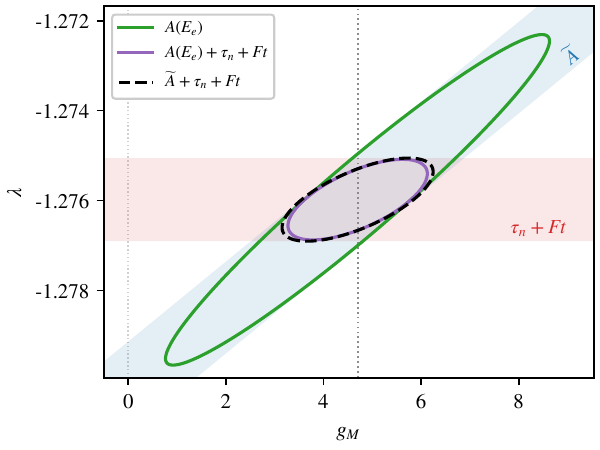}
    \caption{
    Impact of the spectral $A(E_e)$ information on the extraction of weak magnetism, from Asimov datasets mimicking the PERKEO\,III Fierz analysis~\cite{Saul:2019qnp}, i.e., the $[197,694]$~keV window. 
    All regions correspond to $\Delta\chi^2=2.3$.}
    \label{fig:comparacion_fit_lifetime}
\end{figure}

As no public data for $A(E_e)$ are available, we quantify the potential gain with expected-count (Asimov) datasets mimicking two PERKEO\,III analyses~\cite{Markisch:2018ndu,Saul:2019qnp}. 
Counting errors reproduce the published statistical precision, and the remaining uncertainty is represented by a common normalization nuisance parameter. 
Including the independent $\tau_n$ + $\mathcal Ft$ constraint,
the full-spectrum fit reduces $\sigma(g_M)$ by $3.2\%$ in the
$[300,700]\keV$ asymmetry window~\cite{Markisch:2018ndu} and $6.8\%$ in the
$[197,694]\keV$ window of the Fierz analysis~\cite{Saul:2019qnp} compared with a fit to $\tilde A$ alone, cf.~Fig.~\ref{fig:comparacion_fit_lifetime}. At present precision, $\tilde A$ thus captures nearly all the available information on $g_M$.

Further details of this study are given in Appendix~\ref{app:shapeANDvalidation}, where we also use the same numerical approach for three additional purposes. First, we show that the $\tilde A$ prescription, formally derived for energy-integrated measurements, remains an excellent approximation for binned ones. Second, we validate our numerical approach by using it to extract the non-standard Fierz term, finding good agreement with the uncertainties and correlation reported by PERKEO\,III~\mbox{\cite{Saul:2019qnp}}, despite the
limitations of our treatment (e.g., the absence of energy-dependent systematic uncertainties). Finally, we reinterpret that analysis under the SM assumption to extract $g_M$, obtaining a value consistent with our baseline result.

%%%%%%%%%%%%%%%%%%%%%%%%%%%%%%%%%%%%%%%%%%
\section{Conclusions}
\label{sec:conclusions}
%%%%%%%%%%%%%%%%%%%%%%%%%%%%%%%%%%%%%%%%%%

Almost seventy years after Gell-Mann proposed weak magnetism as a hallmark CVC prediction, we report its first observation in the decay of the free neutron, the simplest, nuclear-structure-free $\beta$-decay system.  
Combining the superallowed $\mathcal{F}t$ input, the neutron lifetime, and the neutron $\beta$ asymmetry, we obtain $g_M = 4.47(87)$, which excludes the absence of weak magnetism at $5.1\sigma$ and agrees with the CVC prediction. 
The extraction is robust against the known internal tensions of the neutron dataset: the significance ranges from $5.0\sigma$ to $6.1\sigma$ across all considered variations of the inputs, and it increases when the analysis is restricted to the modern, mutually consistent measurements. We have also shown that, at present precision, the energy dependence of the $\beta$ asymmetry adds little information on $g_M$.

The uncertainty of our determination receives comparable contributions from the three inputs, so progress on any front (next-generation measurements of the neutron $\beta$ asymmetry and lifetime, and improved calculations of the nuclear-structure corrections to the superallowed dataset) will turn this first observation into a precision SM test.
All three fronts are the subject of active, ongoing efforts~\cite{Cirigliano:2019wao,Alarcon:2023gfu,Gorchtein:2023naa,Abele:2025fxh,King:2026vjs}, making such progress a realistic near-term expectation. 
 Together with the determination reported here, these prospects reaffirm $\beta$ decay's standing as a precision laboratory for fundamental interactions.

\begin{acknowledgments}
 We thank Adam Falkowski and Oscar Naviliat-Cuncic for useful discussions.
  We acknowledge the financial support  from the Spanish Ministry of Science, Innovation and Universities (MICIU) and the Spanish State Research Agency (Agencia Estatal de Investigaci\'on MICIU/AEI/10.13039/501100011033) through the  Severo Ochoa Centers of Excellence Programme under Grants No. CEX2024-001441-S and No.~CEX2023-001292-S, the FPU fellowship No. FPU24/01080 and Grant No. PID2023-146220NB-I00. 
\end{acknowledgments}

\newpage
%%%%%%%%%%%%%%%%%%%%%%%%%%%%%%%%%%%%%%%%%%
\appendix
%%%%%%%%%%%%%%%%%%%%%%%%%%%%%%%%%%%%%%%%%%
\clearpage

\section{The effective asymmetry $\tilde{A}$}
\label{app:formulas}

Our starting point is the electron distribution of Eq.~\eqref{eq:dW_recoil}. At first order in recoil, and in the notation of Ref.~\cite{Falkowski:2021vdg}, it reads
\begin{equation}
 \frac{d\Gamma}{dE_e\,d\!\cos\theta}
 \propto w\,\Big[1+\xi_b+\beta P\left(A_{\rm LO}+\Delta A\right)\cos\theta\Big],
 \label{eq:dGamma}
\end{equation}
where $P$ is the neutron polarization, $\beta\equiv p_e/E_e$, and $w$, $\xi_b$ and $\Delta A$ are functions of $E_e$, given by
\begin{align}
 w&=F(Z,E_e)\,p_eE_e(E_0-E_e)^2,
 \label{eq:w}\\[2pt]
 \xi_b&=\frac{2\lambda g_M}{m_N(1+3\lambda^2)}\left(E_0-2E_e+\frac{m_e^2}{E_e}\right)\,,
 \label{eq:xi_full}\\[2pt]
 \Delta A&=\frac{2g_M}{3m_N(1+3\lambda^2)}\big((5\lambda+1)E_e-(2\lambda+1)E_0\big)\,,
 \label{eq:deltaA_full}
\end{align}
where $E_0$ is the electron endpoint (total) energy and $F(Z,E_e)$ is the Fermi function with $Z=1$. The outer radiative corrections and other recoil terms, which are independent of $g_M$ and insensitive to small variations of $\lambda$, have been omitted from Eq.~\eqref{eq:dGamma}, as they play no role in what follows.
 
Let $N_\uparrow$ ($N_\downarrow$) be the number of events with the electron emitted into the hemisphere parallel (antiparallel) to the neutron polarization, within a given energy window. Integrating Eq.~\eqref{eq:dGamma} over $\cos\theta$ and over the window gives the up-down asymmetry, at linear order in recoil:
\begin{equation}
 \frac{N_\uparrow-N_\downarrow}{N_\uparrow+N_\downarrow}
 =\frac{P}{2}\langle\beta\rangle\Big[A_{\rm LO}(\lambda)+\langle\Delta A\rangle_p-A_{\rm LO}(\lambda)\langle\xi_b\rangle\Big],
 \label{eq:updown}
\end{equation}
where
\begin{equation}
 \langle X\rangle\equiv\frac{\int dE_e\,w\,X}{\int dE_e\,w}\,,\qquad
\langle X\rangle_p\equiv\frac{\langle\beta X\rangle}{\langle\beta\rangle}\,,
 \label{eq:averages}
\end{equation}
with the integrals taken over the energy window. The angular term carries the extra velocity factor $\beta$, whereas the isotropic normalization does not. For this reason, $\Delta A$ and $\xi_b$ are averaged with different weights.
 
The experimental collaborations typically subtract the recoil terms in Eq.~\eqref{eq:updown}, evaluated with $g_M=g_M^{\rm CVC}$, and quote the result as $A_{\rm LO}$. The quantity they actually determine is therefore
\begin{equation}
\tilde A\equiv A_{\rm LO}(\lambda)+\langle\Delta A\rangle_p^{\rm res}-A_{\rm LO}(\lambda)\langle\xi_b\rangle^{\rm res}~,
\label{eq:Atilde_def}
\end{equation}
where the superscript ${\rm res}$ denotes the residual left after the CVC subtraction, $X^{\rm res}\equiv X(g_M)-X(g_M^{\rm CVC})$~\footnote{
Over the full kinematic range, $\langle\xi_b\rangle^{\rm res}$ vanishes up to electromagnetic corrections, as WM does not contribute to the total decay rate at this order~\cite{Weinberg:1959zza,Wilkinson:1982hu}. However, this cancellation does not hold for the restricted windows of the actual measurements. }. Expanding around $\lambda_0$, we then obtain Eq.~\eqref{eq:Atilde_linear}:
\begin{equation}
 \frac{\delta\tilde A}{\tilde A_0}=c_A^{\lambda}\,\delta\lambda+c_A^{\rm wm}\,\delta g_M\,,
 \label{eq:Atilde_linear_app}
\end{equation}
where $\tilde A_0\equiv A_{\rm LO}(\lambda_0)$, and
\begin{gather}
 c_A^{\lambda}=\frac{A'_{\rm LO}}{A_{\rm LO}}
 =\frac{(1-\lambda_0)(1+3\lambda_0)}{\lambda_0(1+\lambda_0)(1+3\lambda_0^2)}=-3.11\,,
 \label{eq:clambda}\\[2pt]
 m_N\,c_A^{\rm wm}=\frac{c_A^{\lambda}}{3}E_0+\alpha_1\langle E_e\rangle_p-\alpha_2\Big\langle 2E_e-\frac{m_e^2}{E_e}\Big\rangle,
 \label{eq:cwm}
\end{gather}
where $\alpha_1\equiv-(5\lambda_0+1)/[3\lambda_0(1+\lambda_0)]=5.09$, and $\alpha_2\equiv-2\lambda_0/(1+3\lambda_0^2)=0.434$. The numerical values of $c_A^{\rm wm}$ for the energy windows considered in this work are given in Table~\ref{tab:beta_decay_A0_moments_cAi_summary}.

We note that this prescription is strictly valid for counting asymmetries where a single bin is used. 
However, experimental collaborations often fit energy-binned asymmetries to extract a single value  of $\tilde A$, whose effective WM sensitivity depends on the specific binning. In the following appendix, we evaluate this effect and find that the single-bin approximation underestimates $c_A^{\rm wm}$ by less than $1\%$.

For the remaining correlation observables $a$, $B$, and $\lambda_{AB}$, the appropriate treatment depends on their experimental extraction and sensitivity to weak magnetism.
For the aCORN {\it wishbone asymmetry} measurement~\cite{Wietfeldt:2024acorn}, 
we use the single-bin prescription,  neglecting the detector's momentum-space
acceptance.
As for $\tilde A$, the angular term carries an additional $\beta$ weight, giving
\begin{align}
 \frac{\delta\tilde a}{\tilde a_0}
 &=c_a^\lambda\,\delta\lambda+c_a^{\rm wm}\,\delta g_M,
 \nonumber\\
 c_a^\lambda
 &=\frac{8\lambda_0}
 {(\lambda_0^2-1)(1+3\lambda_0^2)}=-2.76,
 \\
 m_N c_a^{\rm wm}
 &=\frac{1+\lambda_0^2}{2}\,c_a^\lambda E_0
   +\alpha_a\langle E_e\rangle_p
 -\alpha_2
 \left\langle 2E_e\!-\!\frac{m_e^2}{E_e}\right\rangle,\nonumber
 \label{eq:atilde_coeffs}
\end{align}
where $a_{\rm LO}(\lambda)=(1-\lambda^2)/(1+3\lambda^2)$,  $\tilde a_0=a_{\rm LO}(\lambda_0)$,
$\alpha_a=-4\lambda_0/(\lambda_0^2-1)=8.13$,
and $\alpha_2$ and the energy averages are defined above.
For the aCORN energy window $[100,380]\,\mathrm{keV}$, this gives $c_a^{\rm wm}=1.05\times10^{-3}$.
In contrast, for the proton-recoil measurements of $a$~\cite{Beck:2024aspect,Byrne:2002tx,Stratowa:1978gq,Grigoriev:1968zz} and for $B$, whose WM response is negligible, we use the leading-order expressions $a_{\rm LO}$ and $B_{\rm LO}=2\lambda(\lambda-1)/(1+3\lambda^2)$. These measurements therefore
affect the fit
through their constraints on $\lambda$, with $B$ providing only
a minor contribution.
Finally, the WM dependence of
$\lambda_{AB}=(\tilde A-B_{\rm LO})/(\tilde A+B_{\rm LO})$
follows from $\tilde A$, where we assume identical $A$ and $B$ energy windows, the window for $B$ being irrelevant given its negligible WM sensitivity.

\section{Numerical analysis}
\label{app:shapeANDvalidation}

To test the validity of the $\tilde A$ prescription and to quantify the information on $g_M$ that is lost by not exploiting the energy dependence of the asymmetry, we construct neutron $\beta$-decay expected-count (Asimov) datasets~\cite{Cowan:2010js} at $\lambda_0=-1.27598$ and $g_M^{\rm CVC}=4.71$, with a neutron polarization $P=0.991$ as in PERKEO\,III~\cite{Markisch:2018ndu}.  
In these datasets, the bin populations and the two spin counts are set to their expectation values, without statistical fluctuations.
We assign Poisson uncertainties to the spin counts $N_{\uparrow/\downarrow}^i$ in each bin and propagate them to the corresponding up-down asymmetry.

We consider two PERKEO\,III analyses, which use different kinetic-energy windows: the extraction of the asymmetry assuming $b=0$ in Ref.~\cite{Markisch:2018ndu} (hereafter, the asymmetry analysis), with a $[300,700]\keV$ window, and the simultaneous extraction of the asymmetry and the Fierz term $b$ in Ref.~\cite{Saul:2019qnp} (the Fierz analysis), with a $[197,694]\keV$ window. We divide these windows into 31 and 160 equal bins, respectively, following Fig.~4 of Ref.~\cite{Markisch:2018ndu} and the fit binning of Ref.~\cite{Saul:2019qnp} before the factor-of-two rebinning of its Fig.~2.

We also use two calibrations with different numbers of events. In the first, the counting statistics
reproduce the total experimental uncertainty on $\tilde A$ of the
asymmetry and Fierz analyses,  $\sigma_{\tilde A}=(2.1,2.5)\times10^{-4}$, requiring 
$N\simeq(1.3,1.0)\times10^8$ respectively. In the second,
$N\simeq(2.0,2.3)\times10^8$ reproduce only its statistical
component $\sigma_{\tilde A}^{\rm stat}=(1.70,1.65)\times10^{-4}$. 
Although the Fierz analysis excludes about $20\%$ of the data for lack of detector-uniformity measurements~\cite{Saul:2019qnp}, its wider fit window contains more events, as reflected in the statistical calibration.
The remaining systematic uncertainties, $\sigma_{\tilde A}^{\rm syst}=(1.2,1.9)\times10^{-4}$, are represented by a common multiplicative
normalization factor in the predicted asymmetry, profiled with a
Gaussian constraint. Both calibrations reproduce the same total
uncertainty on the fitted  parameter $\lambda$ (or equivalently $A_{\rm{LO}}$). Energy-dependent experimental systematics are not modeled, since the bin-by-bin covariance is unavailable. This is, however, a subdominant effect, as the dominant systematic uncertainties, related to the polarization and magnetic-mirror corrections are normalization-like~\cite{Markisch:2018ndu,Saul:2019qnp}.

\begin{figure}[tb]
 \centering
 \includegraphics[width=\columnwidth]{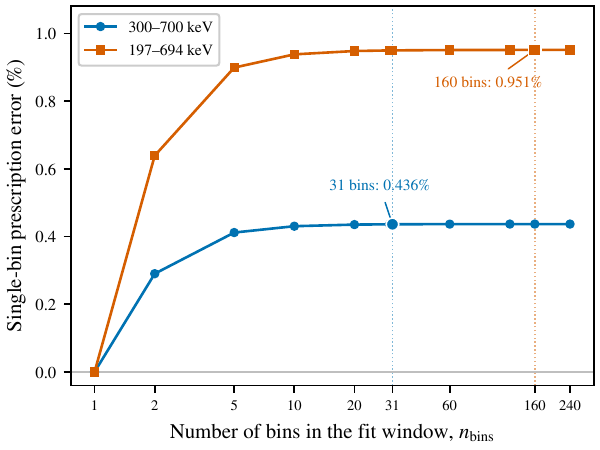}
 \caption{Error in the single-bin approximation to the WM coefficient,
 $1-c_{A}^{\mathrm{wm}}/
 c_{A,\mathrm{eff}}^{\mathrm{wm}}$, in percent as a function of the number of bins in the asymmetry-analysis (blue) and the Fierz-analysis (orange) windows.}
 \label{fig:asimov_single_bin}
\end{figure}

For each Asimov dataset, we fit the binned up-down asymmetries in two ways: with $\lambda$ and $g_M$ as free parameters, and with a single coefficient $A_{\rm LO}$, which we then reinterpret as $\tilde A$. 
To obtain its effective WM coefficient
$c_{A,\mathrm{eff}}^{\mathrm{wm}}$ numerically, we generate Asimov datasets with $g_M$ varied around
$g_M^{\rm CVC}$ at fixed $\lambda_0$ and fit them with a constant $\tilde A$, i.e., assuming $g_M=g_M^{\rm CVC}$ and leaving only $\lambda$ free. 
Repeating this procedure for different bin counts gives
Fig.~\ref{fig:asimov_single_bin}, which is independent of the event calibration. As expected, the
effective coefficient coincides with the single-bin coefficient $c_{A}^{\mathrm{wm}}$
of Appendix~\ref{app:formulas} for one bin, then increases marginally with the number of bins. For 31 and 160 bins, the
single-bin prescription underestimates it by $0.44\%$ and $0.95\%$,
respectively.  Such a correction, to an already recoil-suppressed coefficient, is negligible
at current precision.

In addition to the potential errors induced by the single-bin prescription, the use of a single input $\tilde A$ also removes part of the WM shape information.
We compare the full binned fit with the constant-$\tilde A$ fit
using identical Asimov counts and the independent
lifetime--$\mathcal Ft$ constraint on $\lambda$ in both cases.
When counting errors reproduce the total experimental uncertainty,
the full fit reduces $\sigma(g_M)$ by $2.2\%$ and $3.2\%$ for the
asymmetry and Fierz windows, respectively. With statistically
calibrated counts and the additional normalization uncertainty, the
reductions are $3.2\%$ and $6.8\%$. 
This second treatment reflects the independence of the shape information from normalization errors,  but does not include the energy-dependent detector contributions to the  systematic covariance, which would partly dilute the shape sensitivity.

\begin{figure}[tb]
 \centering
 \includegraphics[width=\columnwidth]{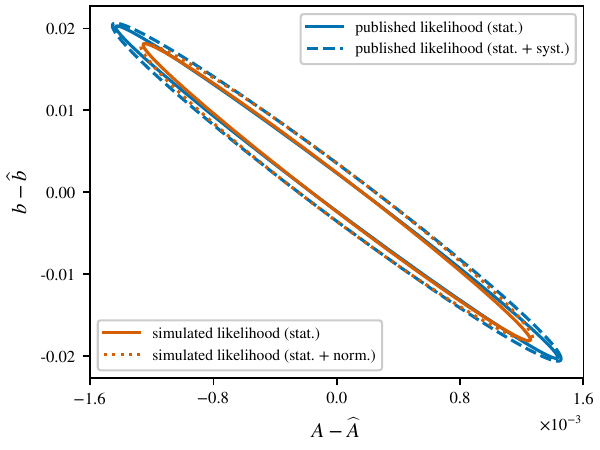}
 \caption{Centered Fierz likelihood contours at $\Delta\chi^2=1$ in the Fierz-analysis window. The published fit~\cite{Saul:2019qnp} is shown in blue while the 160-bin Asimov fit is shown in orange.
 Solid curves include statistics only. Dashed blue includes
 experimental systematics, while dotted orange adds normalization
 uncertainty to the same Asimov dataset.}
 \label{fig:asimov_fierz_comparison}
\end{figure}

As a check of this simplified description, we compare our expected likelihood with the one obtained from real data in the full Fierz analysis~\cite{Saul:2019qnp}, where the Fierz term $b$
modifies the leading asymmetry as $A_{\rm LO}/(1+b\,m_e/E_e)$. For this comparison we use the $[197, 694]\keV$ Asimov dataset with
160 bins and statistically calibrated counts, first without and then
with the common normalization uncertainty. No lifetime/$\mathcal Ft$ constraint enters these fits.
Fig.~\ref{fig:asimov_fierz_comparison} shows the published and
expected likelihood contours, each centered on its own best fit.
The total covariance matrix is available~\cite{Saul:2019qnp}, while the pure statistical one is
reconstructed from the $\Delta\chi^2=1$ ellipses in Fig.~3 of that reference.
The calibration reproduces the $b=0$ conditioned asymmetry uncertainty, and the expected statistical uncertainty on $b$ is only about $10\%$ smaller than the published one, a small difference expected from the energy-dependent detector effects not included in the Asimov dataset.

Although the Fierz and WM responses differ, both are approximately
linear in energy over the fit window, which allows us to recast the
published Fierz likelihood as a WM analysis.
We construct Asimov datasets with $b=0$ at different injected
$(\lambda,\dgm)$ values and fit each of them with free
$(\widehat A,\widehat b)$ at fixed $g_M^{\rm CVC}$. This results in a numerical map that can be inverted to give $(\lambda,\dgm)$ as a function of $(\widehat A,\widehat b)$, the best-fit parameters in the Fierz analysis~\cite{Saul:2019qnp}. The map is approximately linear and easily
inverted. For small parameter variations, its local form is
\begin{equation}
 \dgm\simeq k\widehat b,\qquad
 A_{\rm LO}\simeq\widehat A(1+h\widehat b),
 \label{eq:asimov_recast}
\end{equation}
with $k=144$ and $h=-0.976$.  We have verified that nonlinear effects are negligible in the parameter range relevant for this work.

To quantify the information lost in the recast, we fit the same Asimov dataset at the reference point in two ways: directly in terms of $(\lambda,g_M)$, and in terms of $(\widehat A,\widehat b)$ followed by the recast of Eq.~\eqref{eq:asimov_recast}. Using counting uncertainties only and no lifetime--$\mathcal Ft$ constraint in both cases, the resulting uncertainties on $g_M$ differ by only about $1\%$, showing that the recast retains essentially all the WM information.

\begin{figure}[h]
 \centering
 \includegraphics[width=\columnwidth]{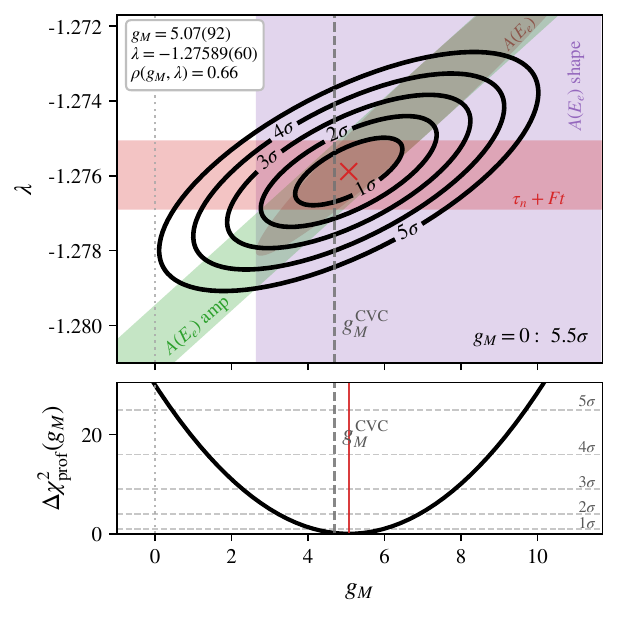}
 \caption{Two-dimensional and profiled constraints on $\lambda$ and $g_M$ from the published Fierz-analysis likelihood~\cite{Saul:2019qnp} obtained through the recast in Eq.~\eqref{eq:asimov_recast}.
 Colored regions separate amplitude (i.e., average), shape and the independent
 lifetime--$\mathcal Ft$ constraint. }
 \label{fig:PerkeoIIIFierz_recast}
\end{figure}

Applying Eq.~\eqref{eq:asimov_recast} to the published Fierz result and
propagating its total statistical-plus-systematic covariance through
the Jacobian of the local map gives
$g_M=7.2(3.0)$, $\lambda=-1.2738(28)$ and
$\rho(\lambda,g_M)=0.972$.
Combining this recast with the lifetime--$\mathcal Ft$
constraint used in the main text yields $g_M=5.07(92)$, excluding
$g_M=0$ at approximately $5.5\sigma$ as shown in Fig.~\ref{fig:PerkeoIIIFierz_recast}. 

To illustrate the different components of this constraint, we separate it into shape and amplitude contributions. To a good approximation, only $\delta g_M$ modifies the spectral shape, while variations of $\lambda$ only change the overall amplitude of the asymmetry, i.e., its average over the energy window. The shape constraint is then obtained by profiling over $\lambda$, which gives $g_M=7.2(3.0)$, whereas the amplitude contribution is the remaining conditional constraint, $\chi^2_{\rm amp}=\chi^2_{\rm recast}-\chi^2_{\rm shape}$.

Since the energy-dependent detector effects, backgrounds and shape systematics may affect the Fierz and WM terms slightly differently, a dedicated experimental WM fit would be desirable. However, as both terms are approximately linear in energy over the fit window, such differences are expected to be small.

\clearpage

\bibliography{main}

\end{document}